\documentclass[aip,rsi,amsmath,twocolumn,superscriptaddress,longbibliography,showpacs,10pt]{revtex4-2}

\usepackage{graphicx}
\usepackage{amsmath}
\usepackage{hyperref}
\usepackage{epsfig}
\usepackage{epstopdf}
\usepackage{amssymb}

\usepackage[ngerman, english]{babel}

\DeclareGraphicsExtensions{.pdf}
\usepackage[utf8]{inputenc}

\usepackage{hyperref}
\hypersetup{%
	colorlinks=true,
	linkcolor=blue,
	urlcolor=blue,
	citecolor=blue
}

\begin{document}

\title{Low temperature and high magnetic field performance of a commercial piezo-actuator probed $via$ laser interferometry}

\author{R. Adhikari}
\email{rajdeep.adhikari@jku.at}
\affiliation{Institut f\"ur Halbleiter-und-Festk\"orperphysik, Johannes Kepler University, Altenbergerstr. 69, A-4040 Linz, Austria}

\author{K. Doesinger}
\affiliation{Institut f\"ur Halbleiter-und-Festk\"orperphysik, Johannes Kepler University, Altenbergerstr. 69, A-4040 Linz, Austria}

\author{P. Lindner}
\affiliation{Institut f\"ur Halbleiter-und-Festk\"orperphysik, Johannes Kepler University, Altenbergerstr. 69, A-4040 Linz, Austria}

\author{B. Faina}
\affiliation{Institut f\"ur Halbleiter-und-Festk\"orperphysik, Johannes Kepler University, Altenbergerstr. 69, A-4040 Linz, Austria}

\author{A. Bonanni}
\email{alberta.bonanni@jku.at}
\affiliation{Institut f\"ur Halbleiter-und-Festk\"orperphysik, Johannes Kepler University, Altenbergerstr. 69, A-4040 Linz, Austria}

\begin{abstract}
	
The advances in the fields of scanning probe microscopy, scanning tunneling spectroscopy, point contact spectroscopy and point contact Andreev reflection spectroscopy to study the properties of conventional and quantum materials at cryogenic conditions have prompted the development of nanopositioners and nanoscanners with enhanced spatial resolution. Piezoelectric-actuator stacks as nanopositioners with working strokes $>100~\mu\mathrm{m}$ and positioning resolution $\sim$(1-10) nm  are desirable for both basic research and industrial applications. However, information on the performance of most commercial piezoelectric-actuators in cryogenic environment and in the presence of magnetic fields in excess of 5\,T is generally not available. In particular, the magnitude, rate and the associated hysteresis of the piezo-displacement at cryogenic temperatures are the most relevant parameters that determine whether a particular piezoelectric-actuator can be used as a nanopositioner. Here, the design and realization of an experimental set-up based on interferometric techniques to characterize a commercial piezoelectric-actuator over a temperature range of $2~\mathrm{K}\leq{T}\leq260~\mathrm{K}$ and magnetic fields up to 6\,T is presented. The studied piezoelectric-actuator has a maximum displacement of $30~\mu\mathrm{m}$ at room temperature for a maximum driving voltage of 75\,V, which reduces to $1.2~\mu\mathrm{m}$ with an absolute hysteresis of $\left(9.1\pm3.3\right)~\mathrm{nm}$ at $T=2\,\mathrm{K}$. The magnetic field is shown to have no substantial effect on the piezo properties of the studied piezoelectric-actuator stack.

\end{abstract}

\date{\today}


\maketitle

\section{Introduction}

The development of precise nanopositioning systems with spatial resolution $\sim$(1-10) nm and time constant $\sim$(10-100) $\mu$s is relevant for both basic and applied research, as well as for industrial applications\cite{Li:2019_SensActA}. The positioning resolution of conventional actuator systems including hydraulic and ac/dc motors is too coarse for most modern technologies, even though these actuators are able to provide large output force and working strokes. The working stroke of an actuator is defined as its linear displacement under dynamic conditions. With the recent development of actuators based on piezoelectric materials $i.e.$ piezoelectric-actuator (PEA), it is now possible to achieve spatial resolution of a few nm\cite{Li:2019_SensActA,Das:2019_RSI}. The application of PEA is wide spread in basic research fields and in industrial sectors, from high resolution scanning probe microscopy (SPM) \cite{Park:1987_RSI,Moore:2008_AnnChem}, to optical systems for astronomy \cite{Koyama:2016_SensA} and aerospace industry\cite{Li:2019_SensActA,Chung:2008_Sens.Act.A,Elahi:2018_Microsyst}. The working stroke of a single piezoelectric element is generally limited to few $\mu\mathrm{m}$ even at room temperature (RT). Alternative approaches have improved the working strokes of hybrid PEA to few centimeters \cite{Wang:2014_IEEE}, but for most commercial PEA, the low working strokes limit their applications. 

A major application area for the PEA is represented by scanning probe measurement systems such as atomic force microscopy (AFM)\cite{Seo:2007_RPP}, scanning tunneling microscopy (STM)\cite{Park:1987_RSI,Moore:2008_AnnChem,Wong:2020_RSI,Song:2010_RSI}, and scanning tunneling spectroscopy (STS)\cite{Zandvliet:2009_AnnChem}. The recent developments in the fields of STM\cite{Park:1987_RSI,Moore:2008_AnnChem}, STS\cite{Zandvliet:2009_AnnChem} and in particular of point contact spectroscopy (PCS) including point contact Andreev reflection spectroscopy (PCAR)\cite{Naidyuk:2005_Book,Naidyuk:2014_SSC,Das:2019_RSI,Daghero:2010_SST,Janson:2012_AJP,Groll:2015_RSI,Andreev:1964_JETP,Andreev:1965_JETP,Soulen:1998_Science}, have underlined the relevance of PEA for nanopositioning applications.  With the emergence of new families of quantum materials encompassing topological insulators\cite{Hasan:2010_RMP}, topological crystalline insulators\cite{Ando:2015_AnnCond}, topological superconductors\cite{Sato:2017_RPP}, Weyl and Dirac semimetals\cite{Armitage:2018_RMP} and unconventional superconductors\cite{Stewart:2017_AdvPhys} including heavy fermionic systems\cite{Stewart:2017_AdvPhys}, PCS and PCAR have proven to be efficient spectroscopic tools to study these material systems\cite{Schwenk:2020_RSI}. Recent investigations of topological crystalline insulators like Pb$_{1-x}$Sn$_{x}$Se and Pb$_{1-x}$Sn$_{x}$Te have pointed at the presence of Majorana fermion-like excitations at the atomic steps of the epitaxial layers \cite{Sessi:2016_Science,Mazur:2019_PRB}. Most PCS and PCAR set-ups reported in literature \cite{Das:2019_RSI,Janson:2012_AJP} are static, with the sample kept fixed while  the tip is the only dynamic component \cite{Janson:2012_AJP,Das:2019_RSI}. Therefore, by introducing a dynamic mode to the static PCAR set-up, a lateral degree of freedom in the sample plane is added and makes it possible to map the sample surface. This improvement is expected to open new perspectives for the characterization of quantum materials. In particular, the use of PEA-based nanopositioning systems along the sample plane promises to facilitate the mapping of surfaces of bulk specimens, thin films and of 2D  layers such as (beyond-) graphene systems \cite{Geng:2018_AdvMater}. The scanning mode PCAR or scanning PCAR (SPCAR) is an exclusive tool for mapping exotic quantum phases and phenomena like Majorana fermions\cite{Alicea:2012_RPP,Elliot:2015_RMP}, weak link Josephson effect\cite{Snyder:2018_PRL}, Meissner and mixed phases of conventional \cite{Hirsch:2012_PhysScr} and unconventional superconductors, odd frequency superconductivity\cite{Linder:2019_RMP} and Yu-Shiba-Rusinov states in magnetically doped superconductors\cite{Scheruebl:2020_NatComm}.

\begin{figure*}[htbp]
	\centering
	\includegraphics[scale=0.65]{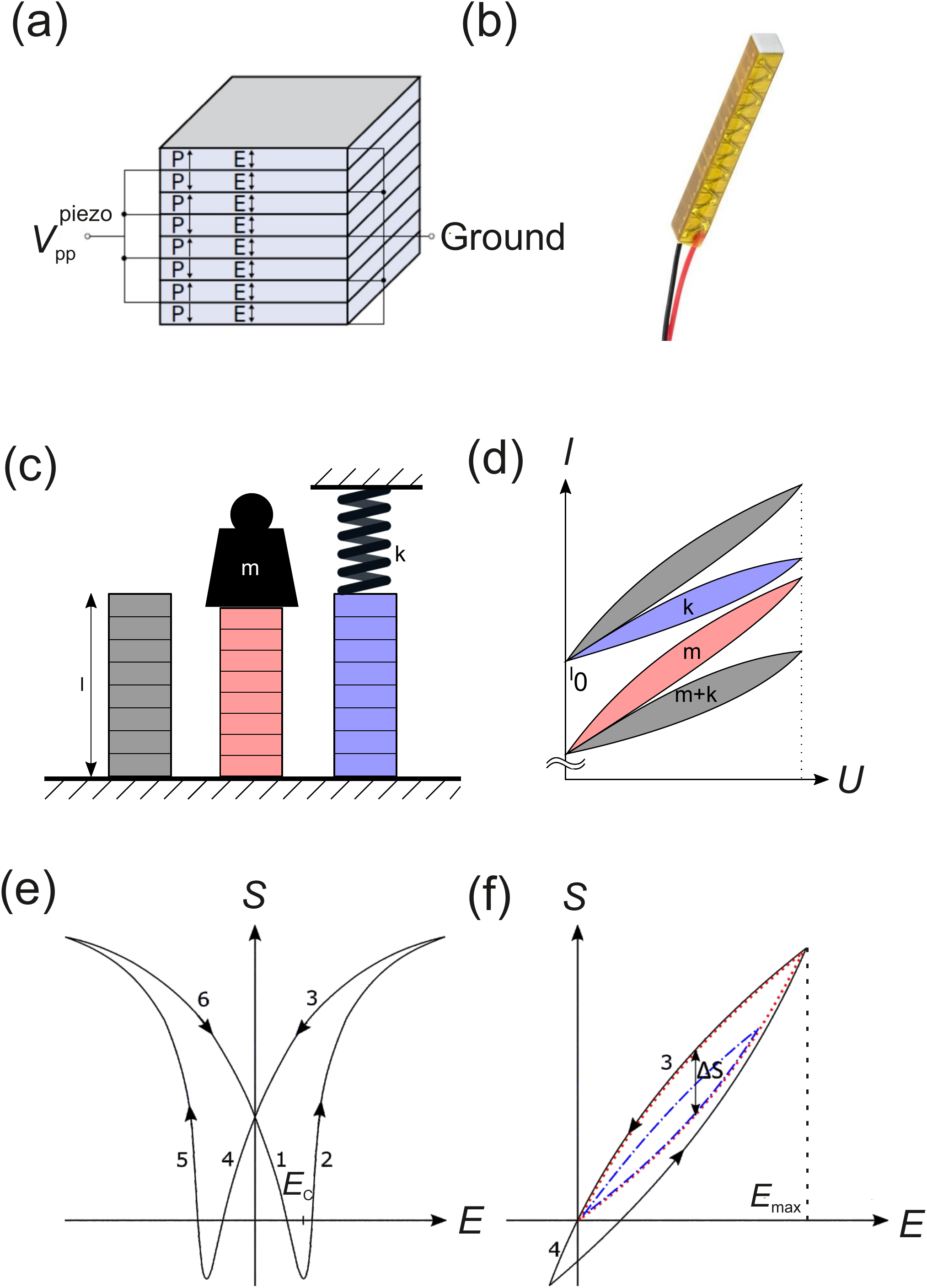}
	\caption{\label{fig:Piezo} (a) Schematic of a PEA stack where multiple PZT layers are mechanically switched in series and electrically in parallel configuration. (b) Commercial PEA stack Model:PK2JUP2 from Thorlabs -- considered in this work. (c) PEA stack: unloaded, mass-loaded and spring-loaded . (d) Hysteresis loops for an unloaded piezo, for a spring-loaded piezo, for a mass-loaded piezo and for a simultaneously spring- and mass-loaded piezo. (e) Strain $S$ in a piezo-element as a function of applied electric field $E$. (f) Operation of a PEA stack when operated under the conditions indicated by the rectangle in Fig.~\ref{fig:fig1}(e).}	
	\label{fig:fig1}
\end{figure*}

The scanning degree of freedom of any physical probe can be accessed by employing either an electromechanical motor or a PEA stack. For applications at cryogenic temperatures, the PEA stacks are preferred over electromechanical motors, since their compactness and physical dimensions favor the integration with state-of-the-art cryostats. In addition, PEA stacks can be controlled by using high precision electronics. However, one major challenge in employing PEA stacks as scanners is represented by their limited displacements at cryogenic temperatures. For most commercial PEA stacks, the absolute displacement reduces with temperature $T$ and  can shrink to  (5-10)\% of the maximum displacement at RT. Most available PEAs have a maximum displacement (20-30) $\mu$m for an applied voltage of (75V-150V) at RT. Moreover, creep and ferroelectric hysteresis associated with a PEA can be detrimental to the application of commercial PEA stacks in cryogenic regime. The piezo-hysteresis as a function of $T$ determines whether a particular PEA stack can be employed as a position scanner with nm resolution, in particular in techniques for mapping surfaces, like STM and SPCAR. Minimal or zero hysteresis of the piezo-scanner is desirable for both spectroscopic and microscopic mapping of surfaces. However, for most commercial PEA stacks, specifications of the PEA stacks at cryogenic conditions are generally not available. 

The absolute displacements of PEAs can be measured by mechanical, electrical or optical means. Due to enhanced signal-to-noise ratio, optical techniques are advantageous over mechanical or electrical sensors such as flexural hinges and strain gauges. In addition, light-based measurements are compatible with cryogenic applications, which can pose challenges to strain  gauge or flexural hinge based sensors \cite{Fleming:2013_SensA}. Optical interferometric techniques have been successfully applied for measurement of nanometer level displacements such as those related to the detection of gravitational waves\cite{Abbott:2009_RPP,Abbott:2018_LivRel}. Here, an indigenously designed and fabricated interferometric set-up for the estimation of displacement, hysteresis and creep of a commercial PEA stack are reported. The measurements have been carried out over the temperature range  $2\,\mathrm{K}<T<260\,\mathrm{K}$, both in the absence and as a function of an applied magnetic field. The absolute displacement of the PEA stack is evaluated to be $25.3~\mu\mathrm{m}$ at RT, while for $T=2\,\mathrm{K}$ an absolute displacement of $1.2~\mu\mathrm{m}$ has been estimated for the maximum allowed voltage of 75 V. Both the displacement and hysteresis of the PEA stack are found to decrease with the $T$ and an absolute residual hysteresis of $\left(9.1\pm3.3\right)\mathrm{nm}$ is estimated at $T=2\,\mathrm{K}$. 

\begin{figure*}[htbp]
	\centering
	\includegraphics[scale=0.75]{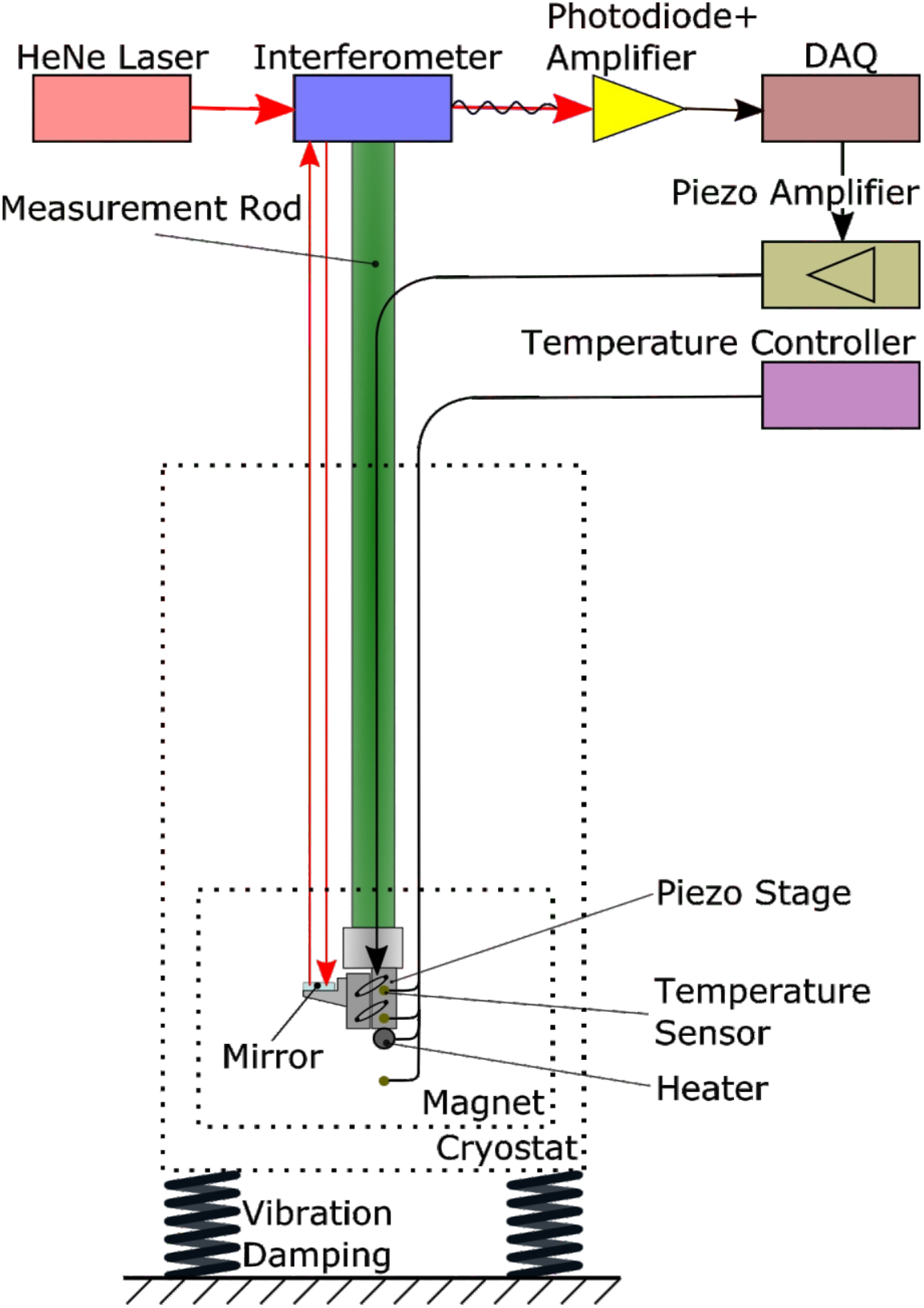}
	\caption{\label{fig:Interferometer} Sketch of the experimental set-up: A piezo-stage inside a cryostat shifts vertically a mirror. The resulting displacement is measured using an interferometer. The whole set-up is vibration damped.  }	
	\label{fig:fig2}
\end{figure*}

\section{Piezo-actuators: Fundamentals}

\subsection{Piezoelectricity: sensors and actuators}

If mechanical stress is applied to a piezoelectric material, an accumulation of opposite electric charges occurs at the opposite faces of the specimen. The effect is related to the formation of electric dipole moments in solids and is known to take place in crystalline systems with inversion asymmetry and in polar materials. The polarization density $\vec{P}$ for such systems is the sum of the dipole moments over the volume of the unit cell. In the piezoelectric effect, the applied mechanical stress leads to a change in $\vec{P}$ and the phenomenon of piezoelectricity depends on the orientation of $\vec{P}$ within the crystal, on the spatial symmetry of the lattice and on the magnitude of the applied mechanical stress. Out of the 21 crystal space groups that are noncentrosymmetric, barring the cubic class, all other 20 crystal classes exhibit direct piezoelectricity. Half of these 20 space group crystals are known to be polar crystals showing spontaneous polarization, \textit{i.e.} for the polar crystals $\vec{P}\ne0$ in the absence of an external mechanical stress. However, for non-polar crystals, the piezoelectric behavior is triggered only upon application of external stress. The application of an external mechanical stress to any piezoelectric material leads to the accumulation of charges, which in turn generates an electric voltage\cite{Uchino:2010_Book}.  

Following the Onsager's reciprocity theorem \cite{Onsager:1931_PhysRev-1,Onsager:1931_PhysRev-2}, the inverse piezoelectric effect in which an applied electrical voltage leads to a mechanical strain also emerges. The existence of an inverse piezoelectric effect was postulated  by Gabriel Lippmann in 1881 and was subsequently demonstrated by the Curie brothers\cite{Uchino:2010_Book}. The first applications of inverse piezoelectric effect were in ultrasonic systems for underwater imaging and communication. The implementation of sensors and actuators based on the piezoelectric effect is widely found in force and acceleration sensors, microphones, 3D printing technology, and medical technology\cite{Uchino:2019_JJAP,Gao:2020_AdvMaterTech}

\subsection{Piezo-actuators: materials}

Most of the current piezo-actuators are produced from donor doped lead zirconium titanate (PZT), which has a ferroelectric transition at $T=623\,\mathrm{K}$. and it is categorized as a soft piezoelectric material \cite{Slouka:2016_Materials}. The  PZTs are generally optimized for: (i) a large coupling factor, which is a parameter defined as the ratio between the mechanical energy generated by the piezo and the electrical energy (due to the applied voltage and the piezo-capacity); (ii) the charge coefficient, $i.e.$ the ratio between the strain developed in the actuator and the applied voltage\cite{Kimura:2017_Book}. By achieving significantly high coupling factors and charge coefficients, the maximum displacement of PZT based actuators can be as high as ~100 $\mu$m at RT.

The schematic of a commercial PEA stack, like the one studied in this work \cite{Piezo_thor}, is shown in Fig.~\ref{fig:fig1}(a). An advantage of employing a stack of PZT elements compared to a single PZT chip is that the maximum displacement is not limited by the material. In the case of the stacked chip architecture of PEAs, every alternate PZT chip, is sintered with silver and connected to every alternate layer forming two electrodes for the application of a voltage. For an $ac$ voltage, the polarization of the PZT is also alternated. A photo of the particular PEA stack used in this work is provided in Fig.~\ref{fig:fig1}(b). 

\begin{figure*}[htbp]
	\centering
	\includegraphics[scale=0.75]{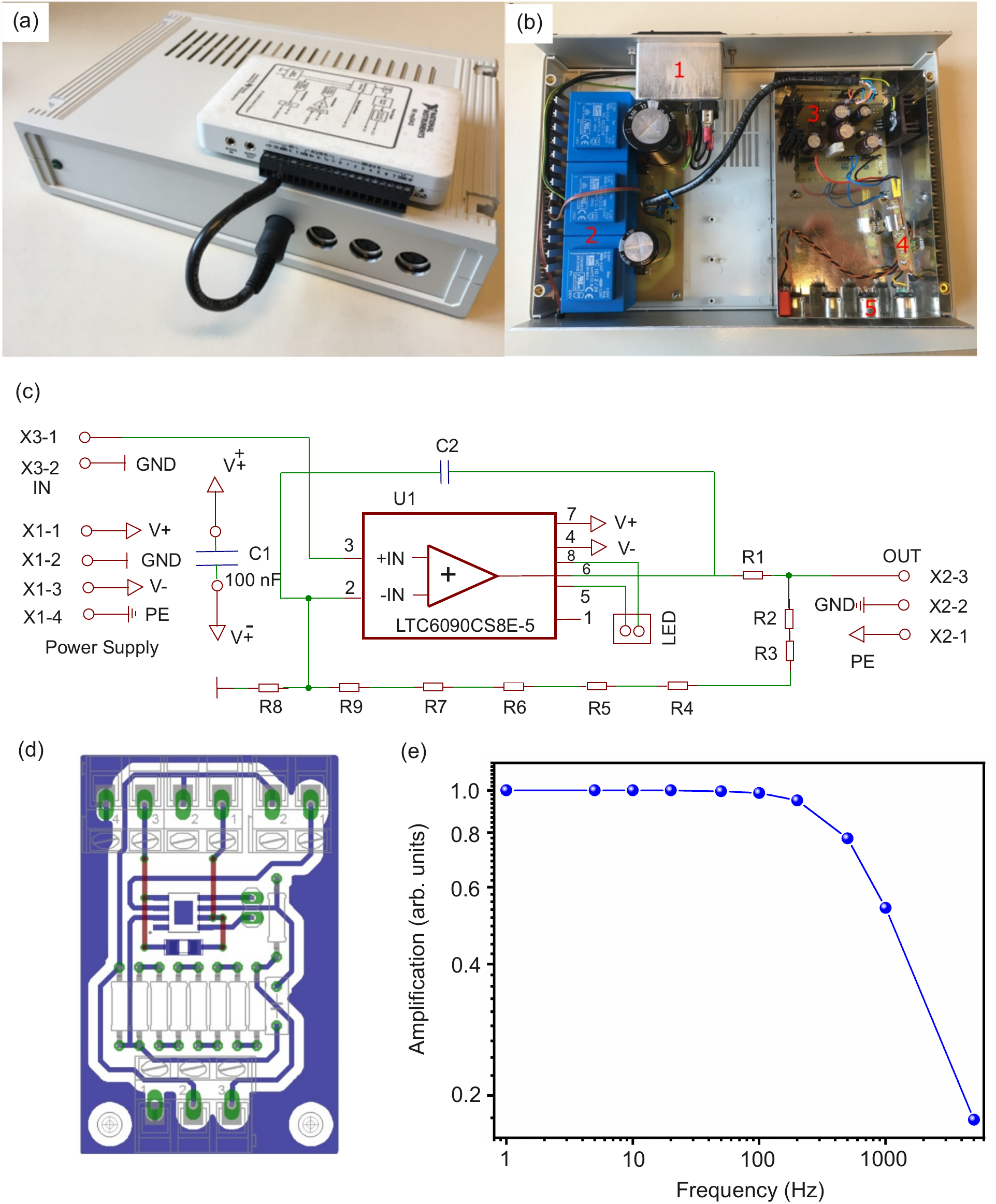}
	\caption{\label{fig:PiezoAmp} (a) Assembled piezo-amplifier driven by a myDAQ. (b) Elements of the piezo-amplifier circuit: (1) mains filter, (2) power supply board, (3) power supply stabilization, (4) amplifier board, and (5) electrical connectors. (c) Circuit diagram of the piezo-amplifier. (d) PCB layout of the piezo-amplifier. (e) Amplification of the piezo-amplifier as a function of the driving frequency.}	
	\label{fig:fig3}
\end{figure*}

\subsection{Piezo-actuators: mechanical and thermal properties}

Most piezoelectric materials are prone to mechanical damage due to the brittle nature of ceramic compounds \cite{Uchino:2010_Book}. In the case of PZT, the ceramic has a compressive strength of $\sim$250 MPa. But at $\sim$10 MPa a mechanical depolarization of the material occurs, affecting the lifetime of the PZT-based actuator. For this reason, the PEA is generally preloaded as little as possible to ensure that the tensile strength of the piezo is not exceeded during dynamic operation. The tensile strength of a material is defined as the stress that the material can withstand while being stretched or pulled before failure. For PZT-based piezo-actuators the tensile strength is estimated to be $\sim$(5-10) MPa. The three categories of preloads generally employed for the dynamic operation of a PEA, namely: (i) mass-loaded, (ii) spring-loaded and (iii) (mass+spring)-loaded are reported in Fig.~\ref{fig:fig1}(c) together with the unloaded PEA stack as reference. A schematic diagram of the corresponding piezo-displacements under the above mentioned preloads is given in Fig.~\ref{fig:fig1}(d). When the PEA is loaded with a mass, it shifts by a constant length, the amplitude of which depends on the load mass and on the piezo-stiffness. This category of PEA displacement is represented by the mass-loaded loop in Fig.~\ref{fig:fig1}(d). For the spring-loaded loop, at $U=0$\,V the spring does not apply any force to the piezo and thus the spring-loaded piezo has the same length $l_{0}$ as the unloaded one. With increasing piezo length, the spring force also increases, resulting in a relatively smaller steepness of the spring-loaded loop in comparison to the unloaded loop. 

\subsection{Piezo-actuators: Hysteresis}

 The hysteresis of the piezo-displacement as a function of the applied electric field is relevant for piezo-actuators and must be considered before choosing it as a nanopositioner or nanoscanner in scanning probe set-ups\cite{Park:1987_RSI,Moore:2008_AnnChem,Seo:2007_RPP,Zandvliet:2009_AnnChem}. The typical hysteresis behavior of a PEA as a function of an applied external electric field $E$ is shown in Fig.~\ref{fig:fig1}(e). The polarization of the PEA is antiparallel w.r.t $E$, resulting in a reduction of the strain with increasing $E$. The electric field is increased from 0 to the critical field $E_{\mathrm{C}}$ along the curve 1. At $E=E_{\mathrm{C}}$, the polarization flips giving rise to a rapidly increasing strain, as indicated by curve 2, until a saturation is reached for $E\gg{E_{\mathrm{C}}}$. For decreasing $E$, the strain decreases until $E=0$ where the polarization is again antiparallel to $E$. This behavior of the PEA is represented by the curves 3 and 4. However, at $E=-E_{\mathrm{C}}$, the polarization flips again and the strain increases and saturates, according to curve 5. A decrease in $E$ results in reduction of the strain until $E=0$, where the segment 6 closes the loop for the entire cycle of $E$. When a piezoelectric material is used as PEA, the flipping of the polarization is generally inhibited by not applying the maximum voltage range allowed for the material. This operation is shown in Fig.~\ref{fig:fig1}(f), which represents a closed loop operation of the segments 3 and 4 in Fig.~\ref{fig:fig1}(e). When the electric field is applied in a cyclic operation from 0 to a maximum $E_{\mathrm{max}}$, a hysteresis is observed in the resulting strain of the piezo-actuator. The hysteresis is defined as:
 \begin{equation}
H =\frac{\left[\Delta{S}\right]_{\frac{E_{\mathrm{max}}}{2}}}{\left[S\right]_{E_\mathrm{max}}}
 \end{equation}
\noindent
where $\Delta{S}$ is the piezo-displacement.
For an unipolar operation of the piezo-actuator, the applied electric field is always positive, as shown by the loop represented by the dotted lines in Fig.~\ref{fig:fig1}(f), while the outermost loop, represented by the solid line, is an example of a bipolar operation. A bipolar operation results in an enhanced displacement range as compared to an unipolar operation, but also in a broader hysteresis, which represents a drawback for applications requiring nanopositioning.

\begin{figure}[htbp]
	\centering
	\includegraphics[scale=0.55]{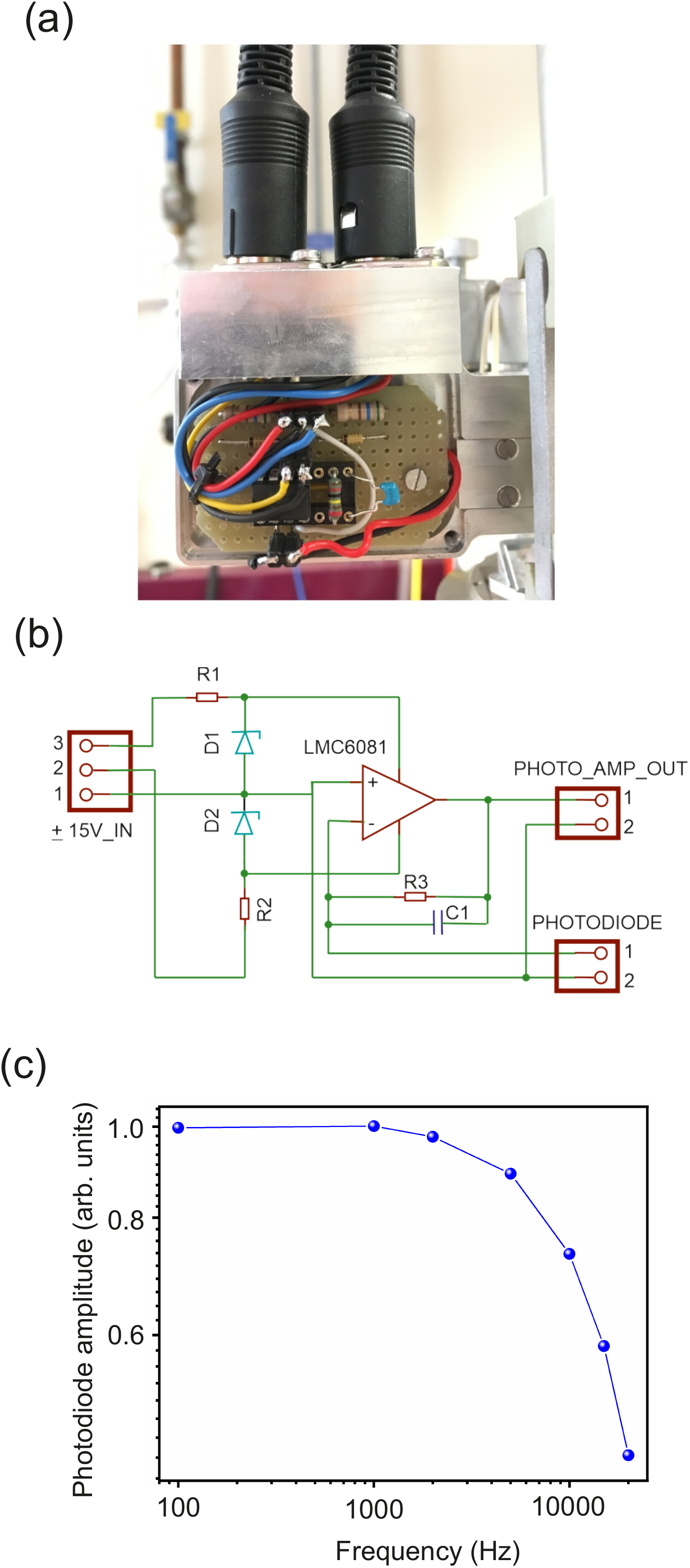}
	\caption{\label{fig:Photodiode} (a) Photodiode-amplifier board in the grounded housing. (b) Circuit diagram of the photodiode-amplifier board. (c) Frequency response of the photodiode-amplifier. }	
	\label{fig:fig4}
\end{figure}

\section{Development of the experimental set-up}

\begin{figure*}[htbp]
	\centering
	\includegraphics[scale=0.50]{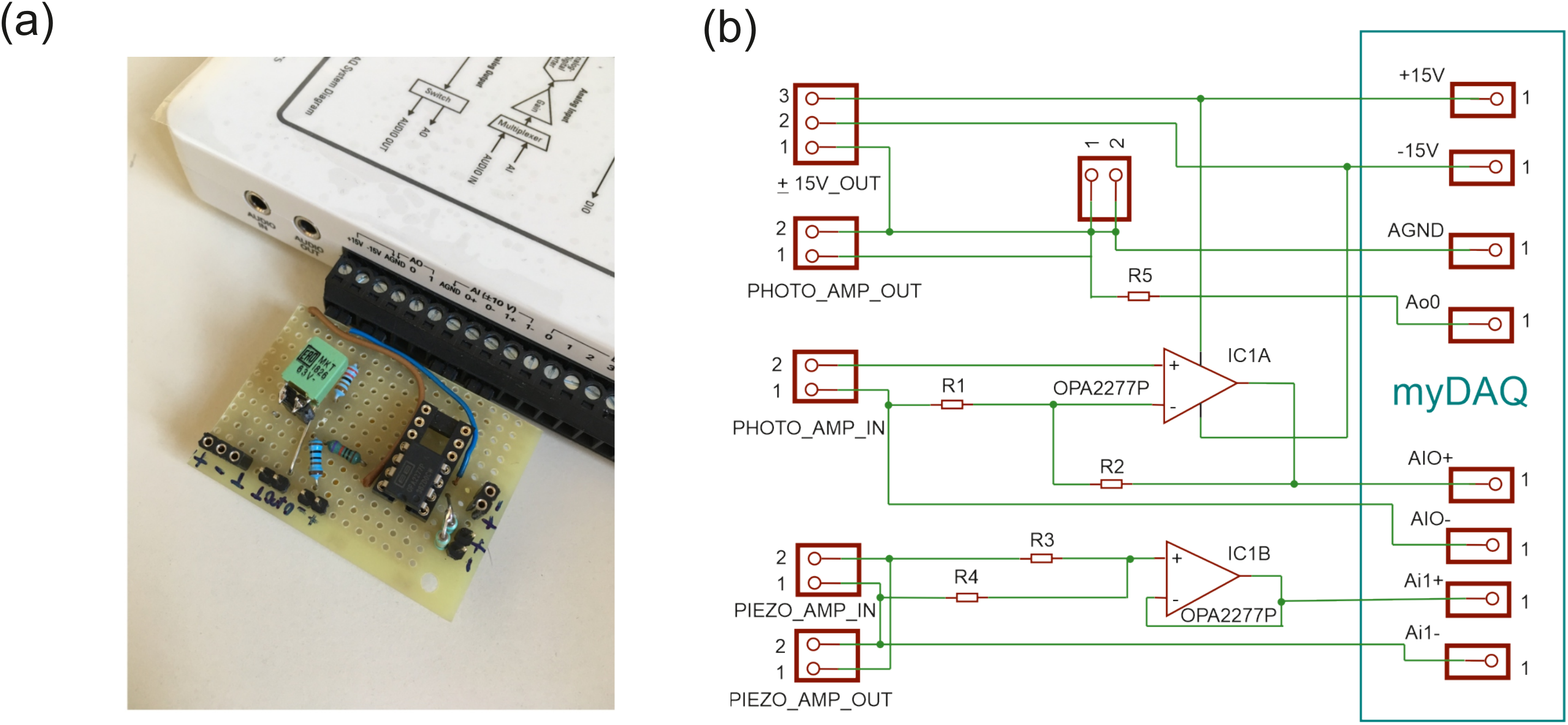}
	\caption{\label{fig:ConnectorBoard}  (a) Connector board and the myDAQ. (b) Circuit diagram of the fabricated connector board.}	
	\label{fig:fig5}
\end{figure*}
An in-depth account of the design and fabrication of the interferometry set-up for measuring the PEA displacement as a function of $T$ and $\mu_{0}H$ is presented in this section. The experimental assembly comprises: (i) the electronics and (ii) the mechanical block. The electronics segment includes the piezo-amplifier, the photodiode-amplifier and the connector board. The components of the mechanical block are: the interferometer, the piezo-stage and the vibration damping assembly.  A schematic representation of the experimental set-up is shown in Fig.~\ref{fig:fig2}. The complete experimental assembly is designed on a 1.0 m long hollow cylindrical sample holder rod (SHR) made out of G10 fiber material. The SHR is designed to fit into a  Janis Super Variable Temperature 7TM-SVM cryostat equipped with a 7\,T superconducting magnet. On one end of the SHR, a mechanical stage is attached which accommodates the PEA on a stage (piezo-stage). The piezo-stage is positioned at the center of the 7\,T superconducting solenoid magnet, as shown in Fig.~\ref{fig:fig2}. On the other end of the SHR, the optical assembly for the Michelson interferometer and for the photodiode-detector is mounted. The SHR connects the interferometer platform to the piezo-stage in a stiff manner. All wires for the electrical connections to the piezo-stage are placed within the hollow SHR. The entire cryostat assembly is placed on an indigenously designed vibration damping mechanism in order to reduce ground vibrations.

\begin{figure*}[htbp]
	\centering
	\includegraphics[scale=1.5]{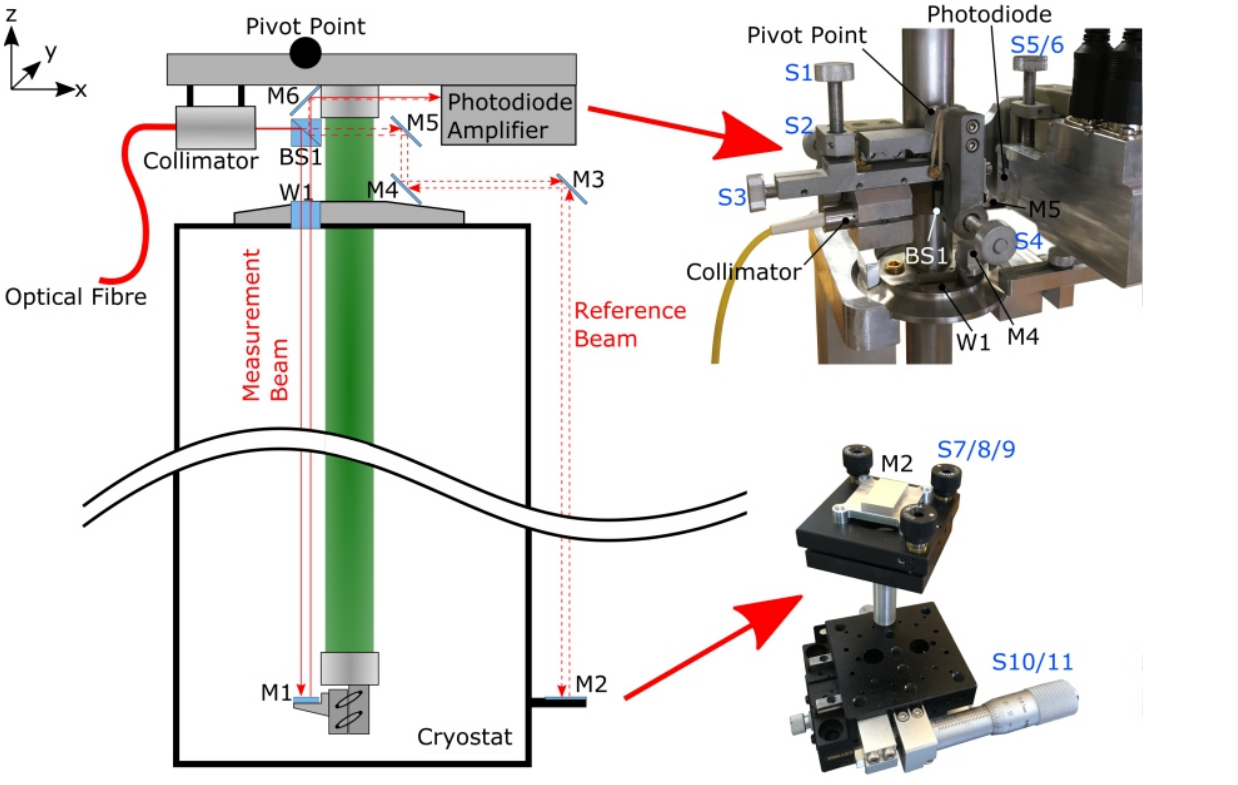}
	\caption{\label{fig:Interferometer2} (left) Schematic of the interferometer inside the cryostat showing the mechanical and optical components. (right) The photographs of the optical parts are also shown. }	
	\label{fig:fig6}
\end{figure*}
 
\subsection{The electronics block}

The electronics block of the experimental set-up consists of a piezo-amplifier used as the power source for the PEA stack, the photodiode-amplifier for detecting the signal from the interference fringes and an electronics connector board. All the electrical and electronic devices designed and used here are controlled \textit{via} a myDAQ from National Instruments. The myDAQ is connected to a computer through an Universal Serial Bus (USB). By using the myDAQ, the displacement of the PEA stack can be estimated from the output voltage of the photodiode-amplifier. The piezo-displacement is controlled by applying an analog \textit{ac} voltage to the piezo-amplifier input.

\subsubsection{The PEA stack}

The PEA stack used in this work is a commercial discrete low voltage PEA stack with a 75\,V of maximum load voltage \cite{Piezo_thor}. The zero load response time of the PEA stack lies in the sub milliseconds range and a free stroke displacement of $30~\mu\mathrm{m}$ at RT. The stack consists of multiple 75\,V maximum voltage based piezoelectric chips fabricated from sintered PZT ceramic. The PEA stack is suitable for open-loop circuit set-ups and for original equipment manufacturers. The stack is equipped with pre-soldered wires and with flat ceramic plates for easy integration into an experimental set-up.

\subsubsection{Piezo-amplifier}

As mentioned in the previous subsection, the PEA stack chosen for this work has an operating voltage range of (0-75)\,V. However, the maximum output voltage from the analog output of the myDAQ is limited to $\pm{10}$\,V. Hence, an amplifier is necessary to exploit the maximum movement range of the  PEA stack. In order to achive a spatial resolution as low as $\sim$1~nm it is essential to employ the maximum allowed voltage for the PEA stack. In this work, the maximum allowed voltage is 75\,V. The main task of the piezo-amplifier is to provide a constant amplification while keeping the noise level of the output $\le25$~mV, which is the minimum voltage required for a piezo-disaplacement $\sim$1~nm. The assembled piezo-amplifier is reported in Fig.~\ref{fig:fig3}. The myDAQ connected with a cable to the main board of the piezo-amplifier is presented in Fig.~\ref{fig:fig3}(a). As shown in Fig.~\ref{fig:fig3}(b), the power cable is plugged directly into the mains filter, labelled as 1 in the figure. The power supply board, marked as 2, converts the \textit{ac} voltage from the mains filter into an unstabilized \textit{dc} voltage. Using the stabilizer 3, the $dc$ voltage is stabilized to provide a low noise supply voltage for the amplifier itself. The amplifier board indicated as 4, is in turn controlled by the input and output connectors labelled as 5. The circuit diagram and the printed circuit board (PCB) layout of the amplifier are shown in Fig.~\ref{fig:fig3}(c) and Fig.~\ref{fig:fig3}(d), respectively. The amplifier realized here is an electrometer amplifier based on the low-noise-high-voltage operational amplifier (OPAMP) LTC6090. By designing the electronic circuit of the amplifier so that the amplifications is stable with $T$, the thermal drifts and fluctuations of the amplifier can be kept at a minimum. The thermal stability of the amplifier is achieved by using eight nearly equal resistors of $\sim$17~$\mathrm{k}\Omega$ as feedback voltage dividers, as shown in Fig.~\ref{fig:fig3}(c). This circuit arrangement ensures that the same power is dissipated by each resistor and that there is no thermal gradient in the feedback voltage divider. Therefore, the ratio between the output and input voltages of the feedback voltage divider are independent of $T$. The purpose of this amplifier is to control the PEA stack, which is a capacitive load. A capacitive load on the output of an OPAMP introduces a negative phase shift, thereby reducing the phase margin and turning the amplifier into an oscillating amplifier. In order to make the circuit stable, a lead compensation (R1,C2) adds a positive phase shift to the curcuit. In the experiments, the PEA stack is driven at frequencies $f<25$~Hz, and therefore no cooling of the OPAMP is required. The frequency response of the piezo-amplifier is reported in Fig.~\ref{fig:fig3}(e), where the amplification of the circuit is measured by sweeping $f$ from 1 Hz to 5000 Hz.  The amplifier bandwidth is estimated to be 600 Hz. Since the PEA stack is operated at $f=25~\mathrm{Hz}$, the measured bandwidth of 600 Hz is sufficient for the application of the PEA stack as nanopositioner and nanoscanner. An increase in bandwidth greater than 600 Hz would augment the noise of the amplifier. A noise level $\sim16~\mu\mathrm{V}_{\mathrm{RMS}}$ has been estimated for the piezo-amplifier. Both the stabilizer and the amplifier are placed inside a grounded tin-plated steel box for shielding against external electromagnetic radiation.

\begin{figure*}[htbp]
	\centering
	\includegraphics[scale=1.5]{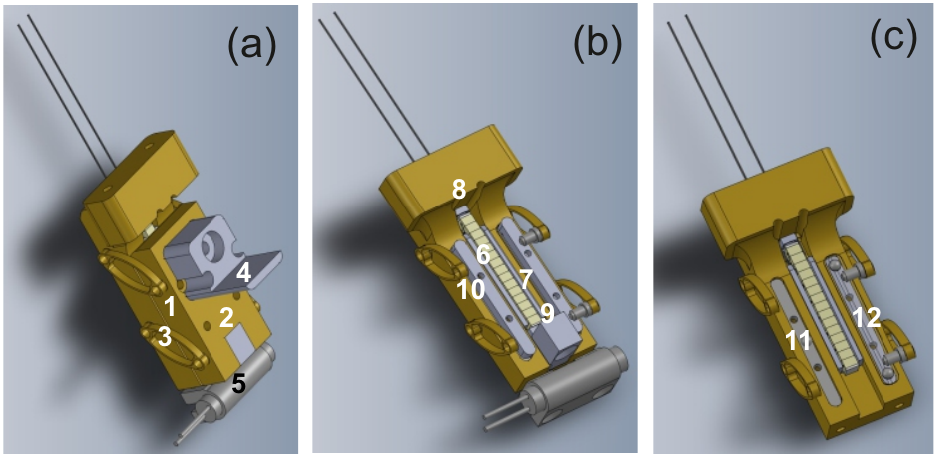}
	\caption{\label{fig:Piezo-stage2}  CAD sketch of the piezo-stage: (a)-(1) base, (2) slide, (3) springs pressing the slide to the base, (4) holder for M1, (5) heater. (b)-(6) piezoelectric-actuator, (7) ceramic case, (8) hemispherical end, (9) tightening block, (10) slide mount. (c)-(11) ball-bearing, (12)  upper and lower tracks.}	
	\label{fig:fig7}
\end{figure*}
\subsubsection{Photodiode-amplifier}

Photodiodes are used as the detectors in light based interefence experiments \cite{Abbott:2009_RPP,Abbott:2018_LivRel} where the interference of two coherent light beams produces concentric and alternate bright and dark fringes. In general, the intensity of light at a particular position of the interference pattern is measured by a photodiode designed to operate at the wavelength of the used light source. In the set-up described here, the photodiode is connected to the amplifier circuit board. The entire photodiode assembly is placed inside an electrically grounded housing and is connected to the connector board through two shielded cables, as shown in Fig.~\ref{fig:fig4}(a). The photodiode-amplifier converts the optical intensity pattern into an electrical signal. The circuit of the photodiode-amplifier is reported in Fig.~\ref{fig:fig4}(b). The photodiode-amplifier is designed based on a single LMC6081 OPAMP which has a low input bias current. The circuit converts the photocurrent of the photodiode into a voltage. The amplification is determined by the resistor R3, while the capacitor C1 ensures stability of the circuit by limiting the bandwidth. Since the maximum supply voltage of LMC6081 is 16 V, the $\pm15~\mathrm{V}$ output voltage from the myDAQ,  must be lowered by the two Zener diodes D1 and D2 and the resistors R1 and R2, as shown in Fig.~\ref{fig:fig4}(b). The frequency response of the photodiode-amplifier is measured by illuminating the photodiode with a modulated light beam. The amplitude of the output signal as a function of the modulating frequency is reported in Fig.~\ref{fig:fig4}(c). The estimated critical frequency of 10 kHz fits well with the chosen values of the circuit components R3 and C1, which are $1~\mathrm{M}\Omega$ and  12 pF, respectively. 

\subsubsection{Connector board}

A connector board serves as the hub between the piezo-amplifier, the photodiode-amplifier and the myDAQ and is shown in Fig.~\ref{fig:fig5}(a). The connector board is directly plugged into the connectors of the myDAQ. The circuit diagram of the fabricated circuit board used in this work is reported in Fig.~\ref{fig:fig5}(b). The $\pm15~\mathrm{V}$ output of the myDAQ is directly connected to the terminal ($\pm15~\mathrm{V}\_\mathrm{OUT}$) for powering the photodiode-amplifier. The output of the photodiode-amplifier is connected to the $\mathrm{PHOTO}\_\mathrm{AMP}\_\mathrm{IN}$ terminal on the connector board. The output signal amplified by a factor of 11 is achieved by using a OPA2277 dual OPAMP in order to make use of the full range of the myDAQ's analog-to-digital-converter (ADC) while maintaining a bandwidth of 100 kHz. The input and output of the piezo-amplifier are connected to the $\mathrm{PIEZO}\_\mathrm{AMP}\_\mathrm{OUT}$ and to the $\mathrm{PIEZO}\_\mathrm{AMP}\_\mathrm{IN}$ terminals, respectively. The output signal of the piezo-amplifier is attenuated by a factor of 11, buffered by the OPAMP and read by the second channel of the ADC. The PEA stack is connected to the $\mathrm{PIEZO}\_\mathrm{OUT}$ terminal, which is switched parallel to the $\mathrm{PIEZO}\_\mathrm{AMP}\_\mathrm{IN}$. The output noise of the myDAQ is further amplified by the piezo-amplifier. The total noise of the electronic system is reduced by a low pass filter in the connector board. The critical frequency of the low pass filter is set to 100 Hz, $i.e.$ far above the excitation frequency of 25 Hz used in this work. Any additional phase shift introduced by the filter does not play a significant role, as the piezo-voltage is directly read using the $\mathrm{PIEZO}\_\mathrm{AMP}\_\mathrm{IN}$ connector. The electronics is controlled by an indigenously developed LabView program and all data are collected $via$ a computer interface.

\subsection{The interferometer}

The detection of a mechanical displacement $\sim$1~nm can be achieved by employing a stable interferometer \cite{Abbott:2018_LivRel,Abbott:2009_RPP}. In this work, a customized Michelson interferometer has been designed and fabricated and the schematic of the set-up is reported in Fig.~\ref{fig:fig6}. The coherent light source used for this experiment is a He-Ne laser with a wavelength of 632.8 nm, which is coupled to a single mode optical fiber. All optical components used in the development of this experimental set-up have been obtained from Thorlabs \cite{Piezo_thor}. As shown in Fig.~\ref{fig:fig6}, the single mode fiber collimates the beam towards the beamsplitter BS1, which splits the incoming laser beam into two beams, each with half the intensity of the original beam. One of the resulting beams is directed downwards into the cryostat through the window W1, fabricated from a N-BK7 glass with anti-reflective coating. This beam is the primary measurement beam. The reference beam is the one which passes through BS1. The primary beam is then reflected by the mirror M1, mounted on the piezo-stage. The reflected primary beam passes through W1 and BS1 and mirror M6 is used to reflect and guide the beam to the photodiode. The reference beam follows the path M5$\rightarrow$M4$\rightarrow$M3$\rightarrow$M2 and it is retroreflected by M2 back to BS1, as shown in Fig.~\ref{fig:fig6}. The beamsplitter BS1 then splits the retroreflected reference beam, which is superimposed with the primary beam. An intereference of the two beams occurs if all the necessary conditions for optical interference are satisfied, resulting in alternate bright and dark circular fringes observed at the photodiode. A condition for intereference is that the path difference between the measurement and the reference beam is shorter than the coherence length of the laser. In order to ensure the conditions for interefence, the mounting height of the mirror M2 -- which is located outside the cryostat -- must be adjustable. The coherence length of the He-Ne laser used here is estimated to be $\sim$150 mm. In this particular experiment, the laser beams travel distances of  several meters. Thus, angular errors of even few tenths of a degree in the optical arrangement can lead to large deviations of the beams, to misalignment and eventually to the disappearence of the interference patterns. For efficient manipulation and alignment of both the primary and reference beams along with the reflected ones, each optical component used in this set-up is designed to be adjustable. The adjustments of the respective optical components are achieved using suitable screws marked by SN with N=1,2,...,11, as reported in Fig.~\ref{fig:fig5}. The collimator, BS1, M5 and M6 are mounted on the same stage which in turn can be rotated freely around a pivot point. The screws S4, S1 and S2 are used to rotate the assembly around the $x-$, $y-$ and $z-$ axes, respectively, while the screw S3 is used exclusively for a linear translation of the stage along the $x-$ axis. The details of this assembly are also shown in Fig.~\ref{fig:fig6}. On the other hand, the module containing the photodiode and the photodiode-amplifier can be manipulated and moved along the $y-$ and the $z-$ axes with S5 and S6, respectively. The screws S7, S8 and S9 are employed to tilt the mirror M2 used for retroreflection of the reference beam, while the screws S10 and S11 serve to manipulate M2 along the $x-$ and the $y-$ axes, respectively. While the mirrors M3 and M4 are fixed, the mirror M1 is moved by the piezo-stage along the $z-$ axis.

The alignment of the interferometer is performed in two steps. First, the primary beam is aligned by adjusting S1, S2, S3 and S4 in such a way that the beam is aligned with the photodiode active area. During this procedure, the reference beam is screened out using a white paper. The mirror M1 is attached to the piezo-stage, placed inside the cryostat, and it is manipulated by rotating the piezo-stage. Once the primary beam is aligned, it is then blocked and the reference beam is unblocked. An iterative approach is taken to tilt and adjust M3 using S10 and S11 until the beam impinges onto M2. Once the beam hits the center of M2, the mirror is tilted and adjusted using S7, S8 and S9 until the reference beam is visible at the photodiode. The primary beam is then unblocked and an interference pattern is observed at the photodiode. The resulting intereference pattern is focused in such a way, that the fringes are broad and defined by fine tuning S7, S8, and S9. It is noted, that the adjustment of M2 does not affect the primary beam alignment.

\begin{figure}[htbp]
	\centering
	\includegraphics[scale=1.0]{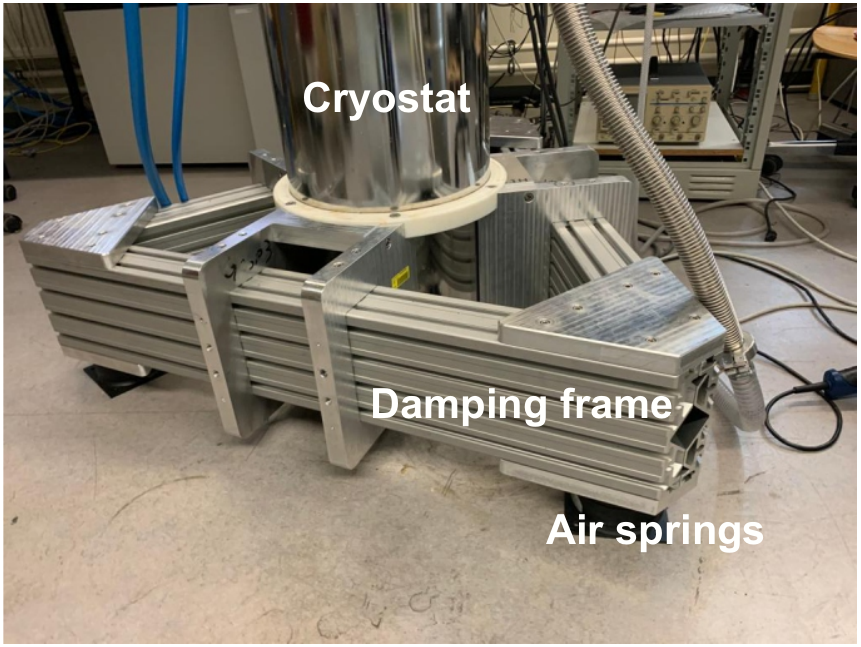}
	\caption{\label{fig:Damping} Damping stage for the cryostat equipped with air springs to compensate ground vibrations.}	
	\label{fig:fig8}
\end{figure}

\subsection{The piezo-stage}

\begin{figure*}[htbp]
	\centering
	\includegraphics[scale=1.25]{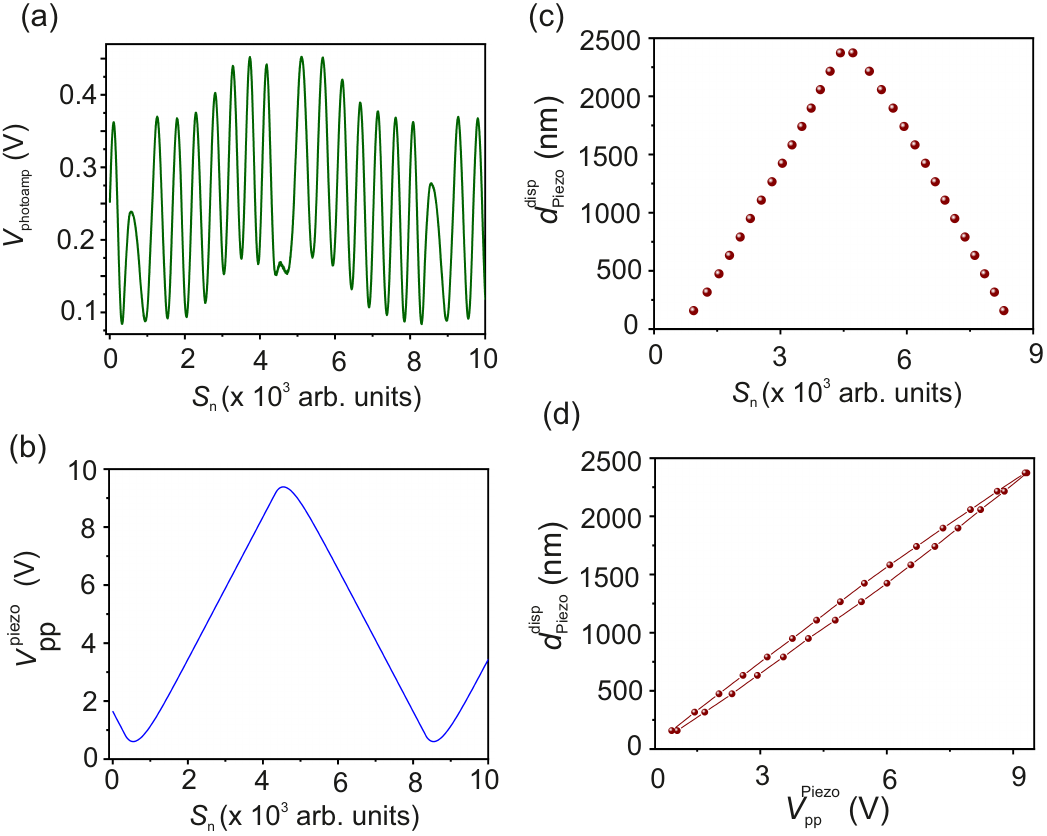}
	\caption{\label{fig:photoampdata} (a) Output voltage of the photodiode-amplifier $V_{\mathrm{Photoamp}}$ as a function of the samplenumber $S_{\mathrm{n}}$.(b) Voltage applied to the PEA stack $V_{\mathrm{Piezo}}$ as a function of $S_{\mathrm{n}}$. (c) Position
		of the PEA stack as a function of $S_{\mathrm{n}}$. (d) Position of the PEA stack as a function of the applied $V_{\mathrm{Piezo}}$ exhibiting a hysteresis loop.}	
	\label{fig:fig9}
\end{figure*}

The piezo-stage serves to transmit the length variation of the PEA stack to the mechanical system which houses M1. The piezo-stage also protects the PEA stack from mechanical failure and ensures that only compressive stress is applied. A computer-aided design (CAD) schematic of the piezo-stage is shown in Fig.~\ref{fig:fig7}. As shown in Fig.~\ref{fig:fig7}(a), the base (1) is the backbone of the whole piezo-stage on which the slide (2) is allowed to move up and down along the $z-$ direction. The whole system is preloaded with four aluminum springs denoted by (3) in Fig.~\ref{fig:fig7}(a). The springs provide lateral and longitudinal forces between the base and the slide. The mirror M1 is mounted $via$ a holder (4). A Ni-Cr cartridge heater (5) is placed at the bottom of the stage which is used for the temperature regulation $via$ a Model-332 Lakeshore temperature controller. The design of the piezo-stage without the slide is shown in Fig.~\ref{fig:fig7}(b). The PEA stack (6) is positioned at the center of the stage and is surrounded by a ceramic case (7), which isolates the piezo-actuator from the conducting base. On the lower side of the piezo-actuator, the tightening block (9) is screwed to the slide prividing tension to the springs while on the topside, the piezo-actuator is fixed against the base through a ceramic hemispherical element (8) that minimizes the bending and the shearing stress. In order to minimize the friction, three ceramic spheres (11) with the ability to roll on molybdenum tracks are placed between the slide and the base. The upper tracks (10) are attached to the slide, while the lower tracks are fixed to the base using molybdenum screws. Two cernox temperature sensors are mounted at the backside of the piezo-stage to measure the temperature at the top and bottom of the piezo-actuator. These two cernox sensors in combination with a permanent cernox sensor located at the bottom of the sample chamber of the  Janis Super Variable Temperature 7TM-SVM
cryostat are used to set and control the temperature at the piezo-stage. The piezo-stage is mounted at the bottom of the SHR and cables are soldered with a cryo-compatible soldering paste to the PEA stack, cernox sensors and the cartridge heater. The cables are led upward through the hollow rod, as already mentioned before.

\begin{figure}[htbp]
	\centering
	\includegraphics[scale=2.0]{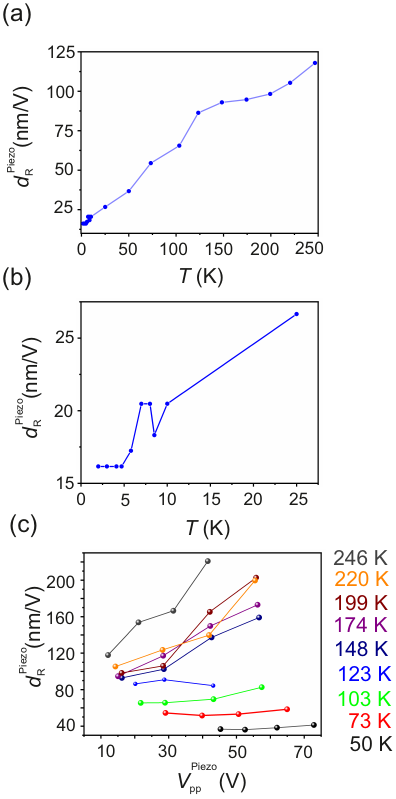}
	\caption{\label{fig:PD} (a) Displacement rate of the PEA stack $d_{\mathrm{P}}$ as a function of $T$ for $2\,\mathrm{K}<T<250\,\mathrm{K}$. (b) $d_{\mathrm{P}}$ as a function of $T$ for $2\,\mathrm{K}<T<25\,\mathrm{K}$. (c) $d_{\mathrm{P}}$ as a function of the applied peak-to-peak voltage $V_{\mathrm{pp}}^{\mathrm{piezo}}$  measured at different $T$.}	
	\label{fig:fig10}
\end{figure}

During dynamic operation, a PEA stack can be subjected to high accelerations and therby tensile stress may be produced. Since common piezoelectric materials, for instance PZT, are prone to tensile stress, for any dynamic operation they must be protected through preloading. The magnitude of the required preload is estimated from the dynamic force $F_{\mathrm{dyn}}$, given by:

 \begin{equation}
F_{\mathrm{dyn}} =2\pi^{2}mf^{2}\Delta{L}
\end{equation}

where $m$ is the mass of the PEA stack and the mounted load, $\Delta{L}$ is the peak-to-peak displacement of the piezo-actuator and $f$ is the frequency of the applied sinusoidal voltage. In the case considered here, since the piezo-actuator is neither fixed to the base nor to the slide, its own weight (5 g) is the mass $m$. Considering a maximum displacement of $\sim3\,\mu{\mathrm{m}}$ at $T\sim4.2\,\mathrm{K}$, and a signal bandwidth $\Delta{f}=100~\mathrm{Hz}$, a dynamic force $F_\mathrm{dyn}=3~\mathrm{mN}$ is calculated for the piezo-actuator considered here. The estimated value of $F_\mathrm{dyn}$ is within the range of the characteristics of the particular PEA stack used in this experiment and thus no preloading is required. The piezo-stage is also kept mechanically backlash-free by eliminating the preloading of the PEA stack.

\subsection{Damping}

The ground vibrations for this experimental set-up have been measured to be $\sim1\,\mu{\mathrm{m}}$, $i.e.$ of the same order of magnitude as the displacement of the PEA stack in cryogenic conditions. In order to achieve an efficient damping it is required that the entire experimental assembly including the cryostat is stiff and decoupled from the ground $via$ soft springs. The stiffness of the set-up is limited by the length of the cryostat and by the dimensions of the sample rod. The damping system used here consists of a heavy triangular frame manufactured from industry grade aluminium profiles. At the three vertices of the triangular frame, air springs are placed, which produce an air suspension for the 300 kg  cryostat, as reported in Fig.~\ref{fig:fig8}. The cryostat is placed at the center of the frame and is suspended a few mm above ground. The air pressure in the springs is adjusted through a pressure gauge. By inflating the air springs so that the cryostat can be lifted a few millimeters above ground, a noise of $\sim$150 nm due to the ground vibrations is measured, which is less than one-sixth of the vibration measured for an undamped system. Thus, the damping against ground vibrations ensures reliable measurements of the piezo-actuator displacement from the interferometric fringes.

\subsection{Data evaluation}

The estimation of the maximum displacement of the PEA stack and its hysteresis as a function of $T$ and applied voltage is accomplished from the data collected at the photodiode-amplifier by measuring the change in intensity of the interference fringes during a dynamic operation. The output voltage of the photodiode-amplifier $V_{\mathrm{Photoamp}}$, as a function of the sample number, $S_{\mathrm{n}}$ (defined as the number of data points accumulated within a defined integration time) and the voltage $V_{\mathrm{pp}}^{\mathrm{piezo}}$ applied to the PEA stack are recorded as a function of $S_{\mathrm{n}}$ and are reported in Fig.~\ref{fig:fig9}(a) and Fig.~\ref{fig:fig9}(b), respectively. Over the upward slope of the triangular $V_{\mathrm{pp}}^{\mathrm{piezo}}$, a sinusoidal $V_{\mathrm{Photoamp}}$ signal with a constant frequency is collected until the turning point at $S_{\mathrm{n}}=4500$. The signal $\sim$0.05V, due to the ambient light is a background offset in Fig.~\ref{fig:fig9}(a).  Each transition from a peak to a valley and $vice~versa$ indicates that the piezo-actuator has travelled a distance of 158.2 nm, $i.e.$ a quarter of the wavelength of the He-Ne laser used here. The position of the PEA stack $d_{\mathrm{Piezo}}^{\mathrm{Pos}}$ at the peaks and valleys of $V_\mathrm{Photoamp}$ is reported in Fig.~\ref{fig:fig9}(c). For a cyclic operation of the PEA stack, where $V_{\mathrm{pp}}^{\mathrm{piezo}}$ is ramped up from +1~V to +9~V and then down to +1~V, a hysteresis in the estimated values of  $d_{\mathrm{Piezo}}^{\mathrm{Pos}}$ is observed, and reported in Fig.~\ref{fig:fig9}(d). 

\begin{figure*}[htbp]
	\centering
	\includegraphics[scale=1.25]{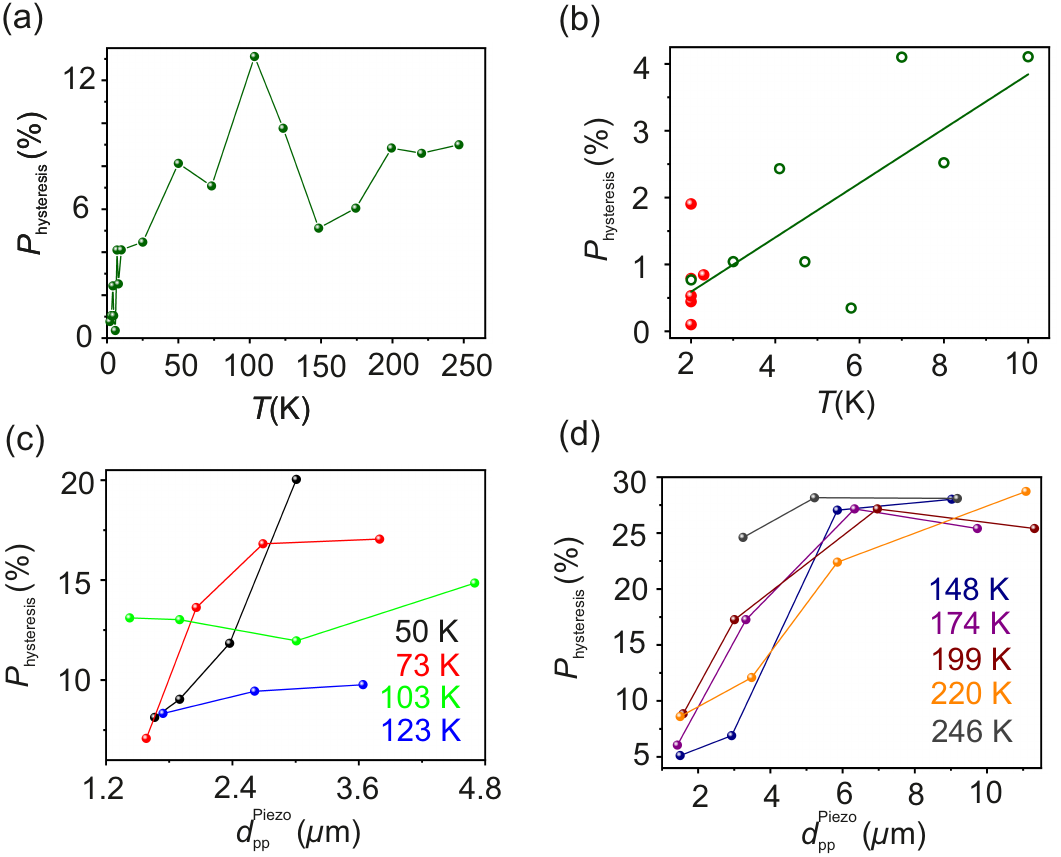}
	\caption{\label{fig:PH}  (a-b) Relative hysteresis of the PEA stack, $P_\mathrm{hysteresis}$ as a function of $T$ estimated for the $T$ range: (a) $2\mathrm{K}<T<250\mathrm{K}$ and (b) $2\mathrm{K}<T<10\mathrm{K}$. (c-d) $P_\mathrm{hysteresis}$ as a function of $V_{\mathrm{pp}}^{\mathrm{piezo}}$ measured for temperatures: (c) 50K, 73K, 103K and 123K. and (d) 148K, 174K, 199K, 220K, and 246K.}	
	\label{fig:fig11}
\end{figure*}

The systematic errors that arise due to disturbances from residual ground vibrations, thermal gradients in the cryostat and turbulances in the liquid nitrogen and helium, are taken into account for the evaluation of the data. One approach is to evaluate multiple samples and to employ statistical methods to reduce the error. The alternative approach follows a method of data fitting leading to an approximation of the real hysteresis loop with an ellipse. This approximation is valid for low $T$ and for low amplitudes of the input voltages due to a decrease of the non-linearity of the PEA stack at these conditions. An ellipse is generally described as a parametric plot of two sinusiodal functions with a phase shift between them. For simplicity, it is assumed that the applied voltage to the PEA stack is a simple sinusoidal function $V(t)$, while the position of the PEA stack is represented by a time dependent function $d(t)$. The excitation frequency of $V(t)$ is $\omega$. The application of $V(t)$ results in a sinusoidal $d(t)$ which is phase shifted by $\phi$ with respect to $V(t)$, due to the hysteresis: 

 \begin{equation}
V(t) = V_{0}  ~\mathrm{sin}\left(\omega{t}\right)
\end{equation}

 \begin{equation}
d(t) = d_{0}  ~\mathrm{sin}\left(\omega{t}\right) + \phi
\end{equation}

When $\phi$ of $d(t)$ with respect to $V(t)$ is known, the absolute hysteresis $H_{\mathrm{abs}}$ and the relative hysteresis $H_{\mathrm{rel}}$ are calculated according to:

\begin{equation}
H_{\mathrm{abs}}=\Delta{d}=d_{0}\mathrm{sin}\left(\phi\right) =d_{0}\phi
\end{equation}

\begin{equation}
H_{\mathrm{rel}}=\phi
\end{equation}

In order to estimate $d(t)$, a representation of the signals in the frequency domain is adopted. When a sinusoidal voltage is applied to the PEA stack, a related voltage can be measured across the photoamplifier. The higher the frequency of the photoamplifier voltage, the higher the absolute value of the speed of the piezo-actuator translation. The mathematical modelling of the data is performed according to the following equation:

\begin{equation}
v_{0}-\left|v\right|=v_{0}-\left | \frac{\partial{d}}{\partial{t}} \right |=v_{0}-v_{1}\left|\mathrm{cos}\left(\omega{t}\right)+\phi\right|
\end{equation}

where $v_{0}$ is an offset, $v$ is the velocity of the PEA stack and $\omega=2\pi{f}$ is the driving frequency of the system.

According to the model employed here, $\phi$ is also the phase shift of the piezo position $d(t)$ with respect to $U(t)$, which in turn is equal to $H_{\mathrm{rel}}$. Each time a measurement is recorded, the $V_{\mathrm{pp}}^{\mathrm{piezo}}$ and the $V_{\mathrm{Piezoamp}}$ are acquired for 10 s with a sample rate of 10000/s with an excitation frequency of 25 Hz. Therefore, the fitting method averages over 250 oscillations.

\section{Results and Discussions}

\subsection{Displacement of the PEA stack}

The displacement rate $d_{\mathrm{R}}^{\mathrm{piezo}}$ of a PEA stack is defined as the change in dimension per unit applied voltage and the knowledge of $d_{\mathrm{R}}^{\mathrm{piezo}}$ for a given piezo-actuator at any arbitrary $T$ is essential for an efficient control of the PEA stack or of the mechanical stage to which it is attached. The  $d_{\mathrm{R}}^{\mathrm{piezo}}$ has been measured for $2\,\mathrm{K}\leq{T}\leq{250\,\mathrm{K}}$ and for a constant displacement of 1.2 $\mu$m and is reported in Fig.~\ref{fig:fig10}(a). The displacement rate is found to increase linearly as a function of $T$. For this work, the temperature range $2\,\mathrm{K}\leq{T}\leq10\,\mathrm{K}$ is of particular interest. The piezo-actuator displacement measured in this temperature range is shown in Fig.~\ref{fig:fig10}(b). For $T\geq5\,\mathrm{K}$, $d_{\mathrm{R}}^{\mathrm{piezo}}$ increases as a function of $T$, with the exception of the value collected at $T=8.5\,\mathrm{K}$. The displacement is found to be constant for $T\leq5\,\mathrm{K}$, with a peak-to-peak value of 1.2 $\mu$m measured for an applied peak-to-peak voltage $V_{\mathrm{pp}}^{\mathrm{piezo}}=75\,\mathrm{V}$. In Fig.~\ref{fig:fig10}(c), the behavior of  $d_{\mathrm{R}}^{\mathrm{piezo}}$  as a function of different applied $V_{\mathrm{pp}}^{\mathrm{piezo}}$ is reported. From Fig.~\ref{fig:fig10}(c), it can be concluded that $d_{\mathrm{R}}^{\mathrm{piezo}}$ increases with increasing $T$. 

\subsection{Hysteresis}

The hysteresis of the PEA stack has been estimated according to the method described above. The relative hystereses $P_\mathrm{Hysteresis}$ of the PEA stack for the temperature ranges $2\,\mathrm{K}<T<250\,\mathrm{K}$ and $2\,\mathrm{K}<T<10\,\mathrm{K}$ are reported in Figs.~\ref{fig:fig11}(a) and ~\ref{fig:fig11}(b), respectively. In particular, multiple measurements have been recorded at $T=2\,\mathrm{K}$ and an absolute hysteresis of $\left(9.1\pm3.3\right)$ nm for a peak-to-peak maximum displacement of $1.2~{\mu}\mathrm{m}$ is estimated. With increasing $T$, a broadening of the hysteresis is observed, as shown in Fig.~\ref{fig:fig11}(b). The solid symbols in Fig.~\ref{fig:fig11}(b) represent the multiple measurements of $P_\mathrm{Hysteresis}$ at $T=2\mathrm{K}$, while the hollow ones are the estimated $P_\mathrm{Hysteresis}$ measured at $T\geq2~\mathrm{K}$. The solid line traces a linear fitting of $P_\mathrm{Hysteresis}$ as a function of $T$. The $P_\mathrm{Hysteresis}$ depends also the peak-to-peak displacement $d_{\mathrm{pp}}^{\mathrm{piezo}}$ of the PEA stack. The behavior of $P_\mathrm{Hysteresis}$ as a function of  $d_{\mathrm{pp}}^{\mathrm{piezo}}$ and measured at $T = 50\,\mathrm{K}, 73\,\mathrm{K}, 103\,\mathrm{K}, ~123\,\mathrm{K}$ and at $T = 148\,\mathrm{K}, 174\,\mathrm{K}, 199\,\mathrm{K}, 220\,\mathrm{K}, \mathrm{and} ~246\,\mathrm{K}$ is reported in Figs.~\ref{fig:fig11}(c) and ~\ref{fig:fig11}(d) respectively. For the temperature range  $50\,\mathrm{K}\leq{T}\leq246\,\mathrm{K}$, $P_\mathrm{Hysteresis}$ increases as a function of increasing  $d_{\mathrm{pp}}^{\mathrm{piezo}}$ and saturates for $d_{\mathrm{pp}}^{\mathrm{piezo}}\geq6~{\mu}\mathrm{m}$.

\begin{figure}[htbp]
	\centering
	\includegraphics[scale=1.75]{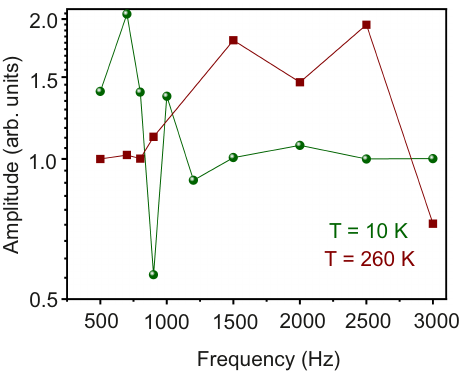}
	\caption{\label{fig:RF} Amplitude of the piezo-stage as a function of frequency at 10~K and 260~K.}	
	\label{fig:fig12}
\end{figure}

\begin{figure*}[htbp]
	\centering
	\includegraphics[scale=0.75]{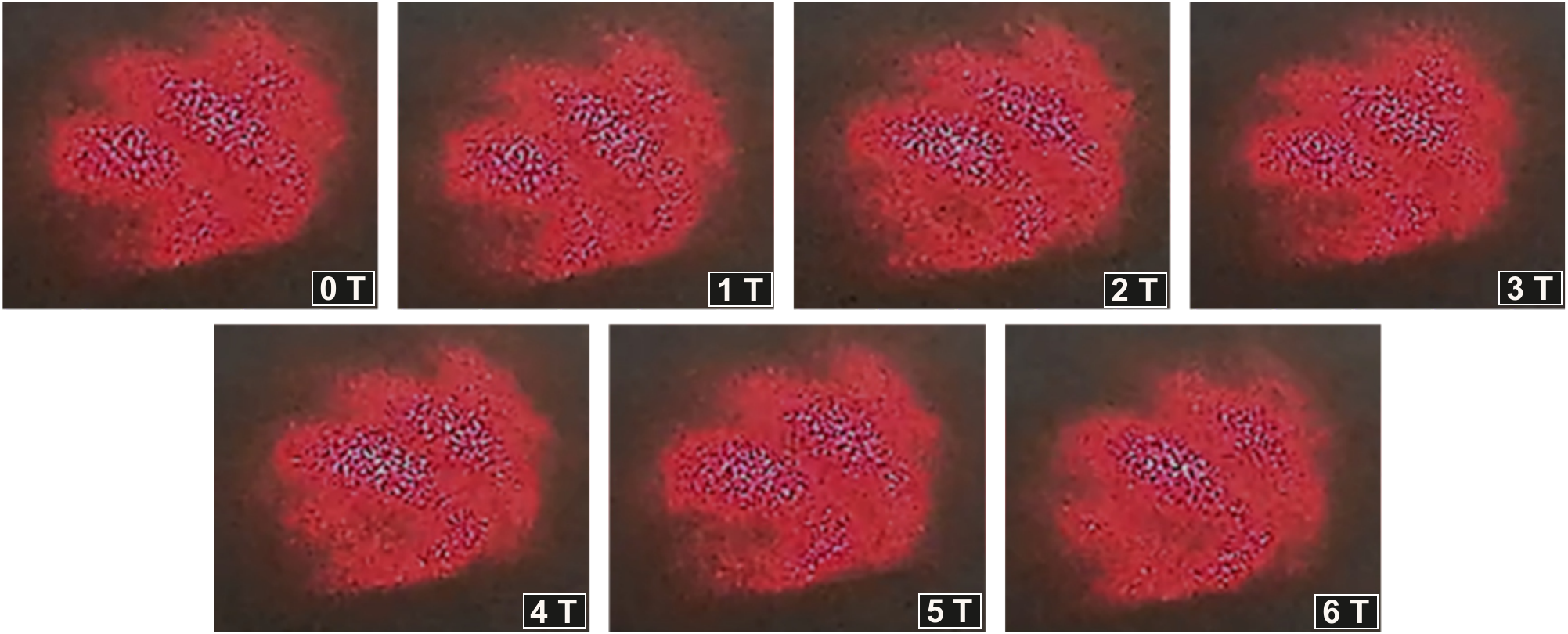}
	\caption{\label{fig:MF} Screenshots of the interference recorded at applied magnetic fields of 0\,T, 1\,T, 2\,T, 3\,T, 4\,T, 5\,T, and 6\,T.}	
	\label{fig:fig13}
\end{figure*}
\subsection{Resonant frequency of the piezo-stage}

The piezo-stage is a mass-spring system in which the mass is a constant quantity. The resonant frequency of the stage is determined by the stiffness of the system during a change of $T$. The frequency response of the piezo-stage has been measured at $T=10\,\mathrm{K}$ and $T=260\,\mathrm{K}$ and is reported in Fig.~\ref{fig:fig12}, where the amplitude of the photodiode-amplifier in arbitrary units is plotted as a function of the driving frequency of the PEA stack. The driving voltage of the PEA stack is kept at $\leq5\,\mathrm{V}$, ensuring that the interferometer works in the linear regime. The correction factors due to the frequency responses of the piezo-amplifier and of the photodiode-amplifier are already taken into account while calculating the amplitude. It is observed, that the resonance frequency of the system changes from $\sim$2500 Hz at $T=260\,\mathrm{K}$ to $\sim$800 Hz at $T=10\,\mathrm{K}$.

\subsection{Effect of magnetic field on the piezo-displacement}

\begin{figure*}[htbp]
	\centering
	\includegraphics[scale=1.25]{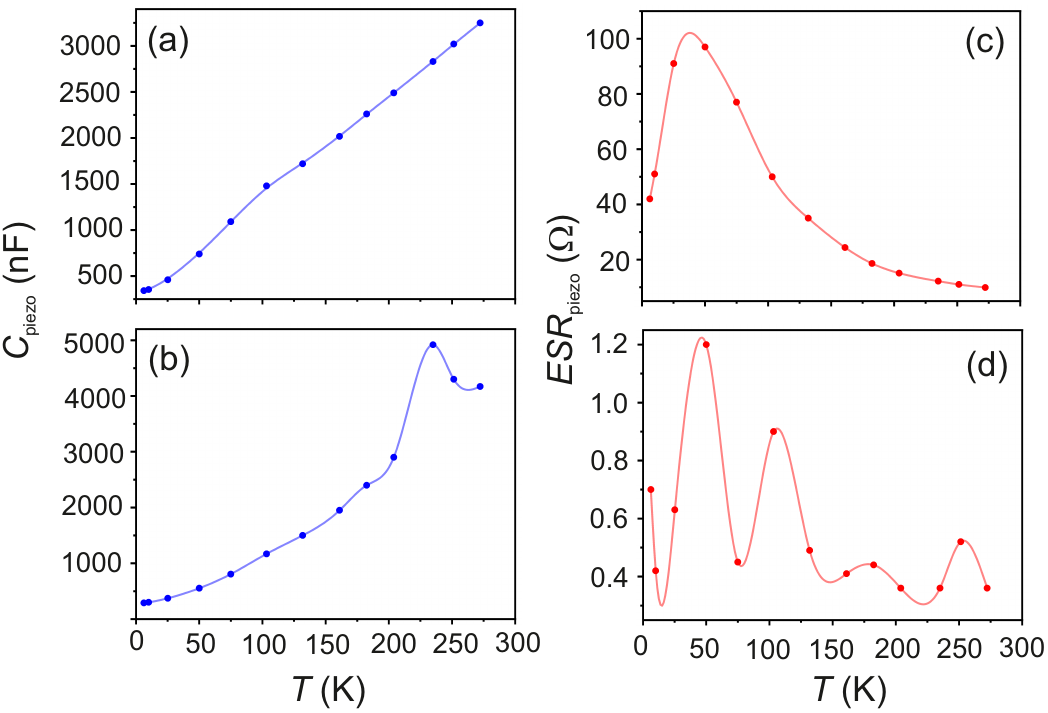}
	\caption{\label{fig:ESR} PEA stack capacity, $C_\mathrm{piezo}$ for applied frequencies of (a) 100 Hz and (b) 100 kHz and ESR measured for applied frequencies of (c) 100 Hz, and (d) 100 kHz over the   range $2\mathrm{K}<T<250\mathrm{K}$.}	
	\label{fig:fig14}
\end{figure*}

The behavior of the PEA stack in the presence of a magnetic field has been studied in real time by monitoring the interference pattern as a function of the magnetic field $\mu_{0}H$, which is ramped from 0\,T to +6\,T. The ramping rate of the magnet is chosen to be 0.01 T/s. The applied piezo-voltage is kept constant during the entire measurement cycle. Any force exerted on the PEA stack or on the piezo-stage due to $\mu_{0}H$ leads to a change in the alignment of the optical path of the measurement and reference beams. Such a misalignment produces changes in the interference pattern or affects the interference conditions. No relevant variation of the diffraction pattern is observed during the sweeping of the magnetic field from 0\,T to +6\,T. A video of the entire measurement cycle is reported in the Supplementary Material \cite{Supplementary}. Screenshots of the interference pattern at $\mu_{0}H=0\,\mathrm{T}, 1\,\mathrm{T}, 2\,\mathrm{T}, 3\,\mathrm{T}, 4\,\mathrm{T}, 5\,\mathrm{T}, \mathrm{and} ~6\,\mathrm{T}$ are shown in Fig.~\ref{fig:fig13}. The slight change in the position of the bright fringes is due to the thermal drift of the set-up. The $d_{\mathrm{pp}}^{\mathrm{piezo}}$ and $P_\mathrm{Hysteresis}$ of the PEA stack for a fixed applied voltage and for a stable $T$ is independent of an applied  $\mu_{0}H$.

\subsection{Capacity and equivalent series resistance}

The capacity $C_\mathrm{Piezo}$ and the equivalent series resistance (ESR) of a PEA stack are crucial parameters for the design of control circuits. A PEA stack can be approximated to an ideal capacitor with an ohmic resistance in series, which in the case of a PEA is the ESR\cite{Uchino:2010_Book}. The estimated $C_\mathrm{Piezo}$ for a driving $f_\mathrm{P} = 100~\mathrm{Hz}$ and $f_\mathrm{P} = 100~\mathrm{kHz}$ as a function of $T$ are given in Figs.~\ref{fig:fig14}(a) and ~\ref{fig:fig14}(b), respectively. The corresponding ESR of the PEA stack measured for $f_\mathrm{P} = 100~\mathrm{Hz}$ and $f_\mathrm{P} = 100~\mathrm{kHz}$ are reported in Figs.~\ref{fig:fig14}(c) and ~\ref{fig:fig14}(d) respectively. For $f_\mathrm{P} = 100~\mathrm{Hz}$, the $C_\mathrm{Piezo}$ increases with increasing $T$, while the ESR has a maximum at $T=50\mathrm{K}$. However, for $f_\mathrm{P}=100~\mathrm{kHz}$, the calculated ESR is lower than the one estimated for $f_\mathrm{P}=100~\mathrm{Hz}$ for all $T$. Thus, the PEA stack studied here is suitable to work at low driving frequencies for applications as high precision nanopositioners and nanoscanners in low-T-high-$\mu_{0}H$ experimental set-ups.

\section{Conclusions}

An interferometry based experimental set-up for measuring the displacement of a commercial PEA stack in the nm range, at temperatures down to 2\,K and under applied magnetic fields up to +6\,T has been designed and realized. The PEA stack is mounted on a piezo-stage and placed inside a cryostat equipped with a superconducting magnet. The displacement of the PEA stack is transferred to the mechanical system, $via$ the piezo-stage equipped with a mirror, which forms the dynamic part of the designed interferometer. The $d_{\mathrm{pp}}^{\mathrm{piezo}}$ and $P_\mathrm{Hysteresis}$ of the PEA stack have been measured as a function of $T$. A monotonous increase of $d_{\mathrm{pp}}^{\mathrm{piezo}}$ as a function of increasing $T$ for $2\,\mathrm{K}\leq{T}\leq250\,\mathrm{K}$ is observed. For $T\leq5\,\mathrm{K}$, a constant $d_{\mathrm{pp}}^{\mathrm{piezo}}=1.2~\mu\mathrm{m}$ is measured for $V_{\mathrm{pp}}^{\mathrm{piezo}}=75~\mathrm{V}$. For $T\leq50\,\mathrm{K}$, a constant $d_\mathrm{P}$ as a function of  $V_{\mathrm{pp}}^{\mathrm{piezo}}$ facilitates an open-loop control of the PEA stack position. At RT, for  $V_{\mathrm{pp}}^{\mathrm{piezo}}=75\,\mathrm{V}$, the measured  $d_{\mathrm{pp}}^{\mathrm{piezo}}=25.3~\mu\mathrm{m}$ measured agrees well with the datasheet provided by the manufacturer. With the decrease of $T$, a reduction in the absolute value of  the $P_\mathrm{Hysteresis}$ is observed. At $T=2\,\mathrm{K}$ and for $d_{\mathrm{pp}}^{\mathrm{piezo}}=1.2\mu~\mathrm{m}$, a residual maximum absolute hysteresis of $\left(9.1\pm3.3\right)~\mathrm{nm}$ is measured for the PEA stack. It is also demonstrated, that the $P_\mathrm{Hysteresis}$ depends on the $d_{\mathrm{pp}}^{\mathrm{piezo}}$ and saturates for $d_{\mathrm{pp}}^{\mathrm{piezo}}\geq6~\mu\mathrm{m}$. In line with the frequency response of the stage, it is concluded that the operating frequency should be kept far below the resonant frequency. Further, an external applied $\mu_{0}H=+6\,\mathrm{T}$ is found to have no effects on neither the piezo-stage nor the PEA stack. Thus, the laser interferometric technique described here can be used for the characterization, over a large range of temperature and magnetic fields, of standard PEA stacks for applications as nanoscanners and nanopositioners in scanning probe techniques and also in astronomy based-, aerodynamics and space technologies.

\section*{Supplementary Material}
A video of the behavior of the interference patterns for a magnetic field sweep from 0\,T to +6\,T is reported in the Supplementary Material \cite{Supplementary}

\section*{Data Availability}

The data that support the findings of this study are available from the corresponding author upon reasonable request and within the article and its Supplementary Material.

\section*{Acknowledgements}

The work was funded by the Austrian Science Fund (FWF) through Projects No. P26830 and No. P31423. The authors thank Prof. David Stifter and Dr. Bettina Heise for the help with the He-Ne laser and the optical fibers used in this work. The authors also thank Prof. Gzregorz Grabecki, Institute of Physics, Polish Academy of Sciences, Warsaw, Poland for the calibration of the cernox sensors used in this work. The authors also acknowledge the technical assistance of Ing. Ekkehard Nusko.

\bibliographystyle{apsrev4-2}

\end{document}